# Experimental observation of bulk Fermi arc in single dielectric resonator

N. Solodovchenko,[1,2,3] F. Zhang,[1,2] M. Bochkarev,[2] K. Samusev,[2,3] M. Song[1], A. Bogdanov,[1,2*] and M. Limonov[2,3]

[1]*Qingdao Innovation and Development Center, Harbin Engineering University, Qingdao, 266000, Shandong, China.*

[2]*School of Physics and Engineering, ITMO University, Saint-Petersburg 197101, Russia*

[3]*Ioffe Institute, St. Petersburg, 194021, Russia*

**Abstract.** The bulk Fermi arc is a fundamental non-Hermitian topological feature that connects two exceptional points (EPs), featuring the transition between Hermitian and non-Hermitian worlds. The bulk Fermi arc emerges when losses are introduced into a Hermitian system, causing a Dirac point to split into two EPs, where both the eigenvalues and eigenfunctions coalesce. Although theoretically predicted in various systems, experimental confirmation has been limited to a two-dimensional photonic crystal slab. Here, we present the first experimental observation of a bulk Fermi arc in a single dielectric resonator. Specifically, we consider a ring resonator made of high-refractive index ceramic. The inner radius and height are varied, enabling the observation of a two-sheeted Riemann surface with two EPs connected by a bulk Fermi arc, confirmed through numerical calculations and experimentally measured extinction spectra at GHz frequencies. These results establish dielectric resonators as a powerful platform for investigating non-Hermitian topological physics and open new avenues for designing topologically robust photonic devices and EP-based sensors.

## 1. INTRODUCTION

Topological photonics emerges as a bridge between the properties of bulk and surface states to create perfect photonic devices using imperfect interfaces [1]. A new page in topological photonics is being opened by non-Hermitian systems that have unique topological hallmarks that are absent in their isolated Hermitian counterparts [2-8]. A striking combination of non-Hermiticity and topology is the "Fermi arc" in photonic structures. Surface Fermi arcs connect ideal *Weyl points*, while bulk Fermi arcs connect *exceptional points* (EPs) [9-13]. The fundamental difference between the two types of Fermi arcs is that in non-Hermitian systems, Fermi arcs represent a bulk phenomenon (similar in this respect to standard Fermi surfaces), whereas, in Hermitian systems, the surface Fermi arcs connect the projection of the Weyl points to a given surface [14,15]. Unlike the traditional view that frequency levels are closed curves, each Fermi arc is an open isofrequency curve ending at two singular points.

Weyl points in periodic systems, corresponding to gapless singularity in the k-space around which bands disperse linearly with respect to three quasimomenta, can be present in various three-dimensional band structures. The first experimental observation of surface Fermi arcs was demonstrated on an ideal Weyl system in which all Weyl points are symmetry-related, exist at the same energy, and are free from nontopological bands. In a 3D photonic crystal composed of metallic inclusions, four Weyl points and helicoidal dispersion were observed, resulting in the open Fermi arcs connecting points of opposite chirality [12]. In the presence of non-Hermiticity, a Dirac point with a nontrivial Berry phase can split into pairs of isolated EPs. The resulting two-sheeted Riemann surface associated with these pairs of EPs, in turn, leads to the formation of a bulk Fermi arc, which connects the two EPs in a complex band structure (see Supplementary Information, section I). At the EP, both eigenstates and eigenvalues of the structure coalesce, unlike diabolic points [16], where only the real part of eigenvalues coalesce. EPs are among the most striking and universal features of non-Hermitian physics [2,6]. Note that in addition to individual EPs, the existence of higher-order EPs [17], exceptional lines [18], exceptional rings [19], exceptional surfaces [20,21], as well as the experimental demonstration of anisotropic EP [22] have been reported.

The existence of topological EPs and bulk Fermi arc in Dirac materials with two distinct quasiparticle lifetimes was predicted theoretically by Kozii and Fu in 1993. It was demonstrated that, rather than touching at the Dirac point, the quasiparticle conduction and valence bands stick on the Fermi arc, which ends at two topological EPs [9]. However, the first experimental observation of a bulk Fermi arc was reported only in 2018 in a two-dimensional periodic photonic crystal, where radiative losses at the top and bottom of a finite-thickness slab facilitated its emergence. [10]. Unlike the surface Fermi arc, which connects Weyl points with opposite charges (±1), this bulk Fermi arc connects a pair of EPs with opposite chiralities (topological half-charges ±1/2), and along the arc, and along the arc, only the real parts of the eigenvalues coincide, while the eigenfunctions differ.

Bulk Fermi arcs were demonstrated theoretically in various structures, indicating this phenomenon's fundamental nature. In a non-Hermitian superconducting system from a conventional superconductor with a ferromagnet lead, the appearance of EPs when an external Zeeman field is applied was theoretically demonstrated, which leads to the emergence of topologically protected and highly tunable bulk Fermi arcs [23]. The interplay of hydrodynamics and elasticity in ordinary, passive soft matter is shown to split Dirac cones into bulk Fermi arcs, pairing exceptional points with opposite half-integer topological charges [24]. The non-Hermitian description in two-dimensional systems of heavy fermions with momentum-dependent hybridization allows the establishment of the existence of bulk Fermi arcs and reveals their connection to the topological EPs [25]. As a result of the analysis of the influence of disorder on the low-energy behavior of two-dimensional tilted Dirac-fermion systems, in which the fermions have two distinct orbitals unrelated by any symmetry, it is shown that in such a disordered phase, the Dirac point is destroyed and replaced by a bulk Fermi arc [26]. Despite this, the Fermi arc has never been experimentally demonstrated in a single open isolated resonator.

In this work, we experimentally and theoretically demonstrate the existence of the bulk Fermi arc in a single compact scattered – a dielectric ring resonator (RR) with a rectangular cross-section made of high-refractive-index ceramic. A region of parametric space of heights and widths of RR containing the Fermi arc was determined from the calculations of the double Riemann sheet. Then, a series of experimental samples with different inner radii and heights were used to cover this region of the parametric space with EPs and bulk Fermi arc. The complex eigenfrequencies were retrieved from the extinction spectra measured for 203 samples at GHz frequencies using the Fano formula combined with the quasi-normal mode expansion. As a result, we obtained the high-accuracy experimental two-sheet Riemann surface and traced bulk Fermi arc, demonstrating fundamental aspects of a photonic modes evolution in the vicinity of EP.

## Results

**Exceptional points in dielectric rings**

The aim of this work is to experimentally observe the bulk Fermi arc in a single dielectric RR, in whose calculated spectra we previously observed quasi-bound states in the continuum (q-BIC) and EPs connected by the bulk Fermi arc [27]. Here, we use the term "q-BIC" since a genuine BIC with an infinite quality factor does not exist in finite structures [28], with a few exceptions [29, 30]. EPs and q-BICs can be observed in a subwavelength resonator due to the interference of Mie modes located in the same radiation channel [31]. For resonators with cylindrical symmetry, the scattering channels corresponding to different azimuthal indices are independent. The Mie modes of an RR can be conditionally divided into two groups, differing in the number of oscillations in the radial and axial directions, which exhibit different eigenfrequency shifts depending on the ratio of the width $W=R_{out}-R_{in}$ to the resonator height $h$ [31]. In [32], it was shown that the interference between two modes of an RR can be controlled by changing the radius of the inner hole $R_{in}/R_{out}$, which allows observing EPs and q-BICs in an RR [27, 33].

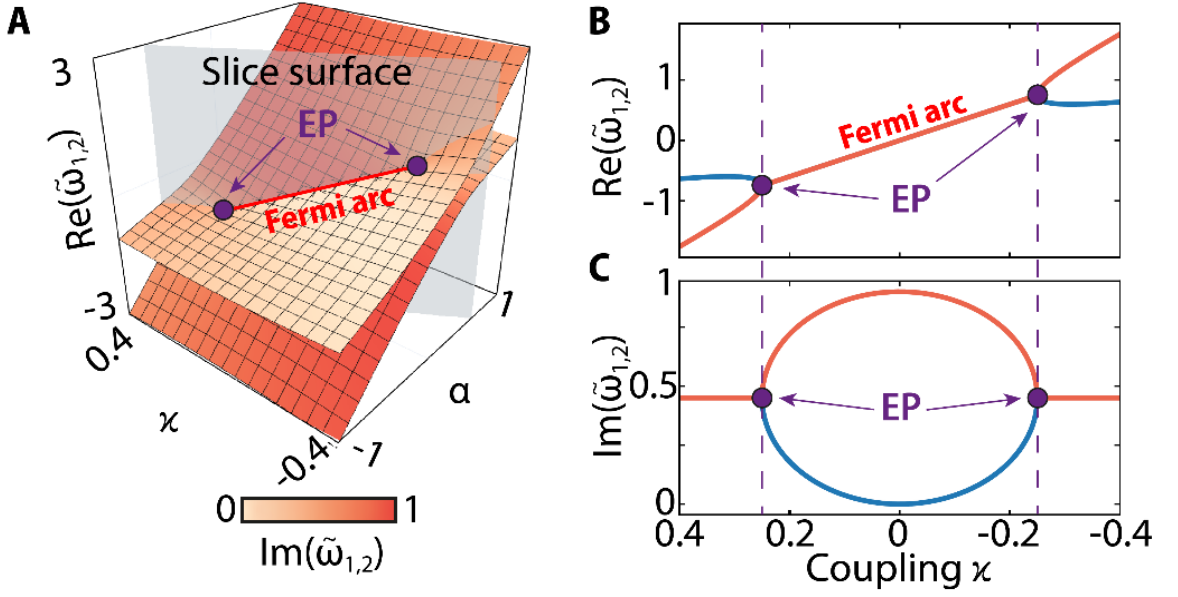

**Fig. 1 | Exceptional points and bulk Fermi arc as a solution of Hamiltonian.** (**A**) Double-Riemann-sheet: real and imaginary (colored) parts of eigenfrequency $\widetilde{\omega}_{1,2}$ of Hamiltonian in parametric space ($\varkappa$, $\alpha$). (**B**) Bulk Fermi arc and (**C**) imaginary part of eigenfrequency $\widetilde{\omega}_{1,2}$ of Hamiltonian relative to the coupling constant $\varkappa$ (projection of the slice surface from (**A**)). The purple dotted lines show EPs positions. Results obtained from the phenomenological effective Hamiltonian (1).

Phenomenologically, the interaction between two modes can be described by the temporal coupled-mode theory (TCMT), and the evolution of an open photonic system is characterized by Hamiltonian [19]:

$$H = \begin{pmatrix} \omega_1 & \varkappa \\ \varkappa & \omega_2 \end{pmatrix} - i \begin{pmatrix} \gamma_1 & \sqrt{\gamma_1\gamma_2} \\ \sqrt{\gamma_1\gamma_2} & \gamma_2 \end{pmatrix}, \quad (1)$$

where $\omega_{1,2}$ and $\gamma_{1,2}$ are resonant frequencies and half-widths of uncoupled resonant states, correspondently; $\varkappa$ is the near-field intrinsic coupling between the states, which operates at the same radiation channel, and their interference produces coupling through the continuum $\sqrt{\gamma_1\gamma_2}$. According to this model, the q-BIC regime occurs when the coupling coefficient $\varkappa$ has the next form:

$$\varkappa_{q-BIC} = \frac{(\omega_1 - \omega_2)\sqrt{\gamma_1\gamma_2}}{(\gamma_1 - \gamma_2)}, \quad (2)$$

and for the appearance of an EP, two conditions have to be satisfied [27]:

$$\begin{cases} \varkappa_{EP} = \pm \frac{|\gamma_1 - \gamma_2|}{2}, & (3a) \\ \varkappa_{EP} = -\frac{(\omega_1 - \omega_2)(\gamma_1 - \gamma_2)}{4\sqrt{\gamma_1\gamma_2}}. & (3b) \end{cases}$$

The effective Hamiltonian (1) determines the two-sheet Riemann surface incorporating two EPs and the Fermi arc. For instance, it can be constructed for fixed values $\gamma_1 = 0.7$, $\gamma_2 = 0.2$ and frequencies $\omega_1 = \alpha$, $\omega_2 = 3\alpha$ with the variable parameter $\alpha$ for the corresponding non-interacting modes, marked with indices 1,2. Figure 1A shows this example, in particular, two EPs appear in the solution of the effective Hamiltonian when conditions from Eqs. (3) hold. These points are connected by a Fermi arc that is observed at $|\varkappa_{EP}| < |\gamma_1 - \gamma_2|/2$. The second condition for an EP, Eq. (3b), can be represented using a "slice surface" in Figure 1A, and its projection demonstrates the Fermi arc dependence on the coupling parameter $\varkappa$ (Fig. 1B and 1C).

**Bulk Fermi arc in dielectric rings: calculations**

We selected a pair of RR's modes corresponding to radial and axial oscillations, for which changes in aspect ratios $W/h$ or $R_{in}/R_{out}$ produce a reduction of the spectral splitting $\Delta\omega$. This also leads to modes of interference and changes in the resonant frequencies and quality factors. The inner hole $R_{in}/R_{out}$ is mainly responsible for changing the coupling parameter $\varkappa$, while $W/h$ is mostly for the relative frequency difference between the modes.

Thus, for RR with $\varepsilon = 80.05$, losses $\tan\delta = 1e{-4}$, and $R_{in}/R_{out} = 0.499$ (Figs. 2B and 2C), by changing the aspect ratio $W/h$, one can achieve either q-BIC (at $W/h \sim 0.439$) with magnetic quadrupole dominance [31] and characteristic field distribution (Fig. 2A2) or EP (at $W/h \sim 0.455$) with two identical eigenvectors (Figs. 2 A3/B3). For EP, the eigenmode fields $|E|$ have identical distributions (Fig. 2A). The bulk Fermi arc connects two singular points at $\varkappa_{EP} = \pm(\gamma_1 - \gamma_2)/2$ (see Supplementary Information, section I). In the case of a single RR, it can be observed by varying two geometrical parameters, for instance, $R_{in}/R_{out}$ or $W/h$. The coupling parameter $\varkappa_{EP}$ can be obtained by fitting using calculated eigenvalues and the effective Hamiltonian (1) [27]. Figures 2D and 2E show surfaces of real and imaginary parts of dimensionless eigenfrequencies $kW$ in the parametric space $(R_{in}/R_{out}, W/h)$. The bulk Fermi arc itself is marked by the red line in Fig. 2D. Along the arc, the real component of eigenfrequencies is the same for both photonic modes.

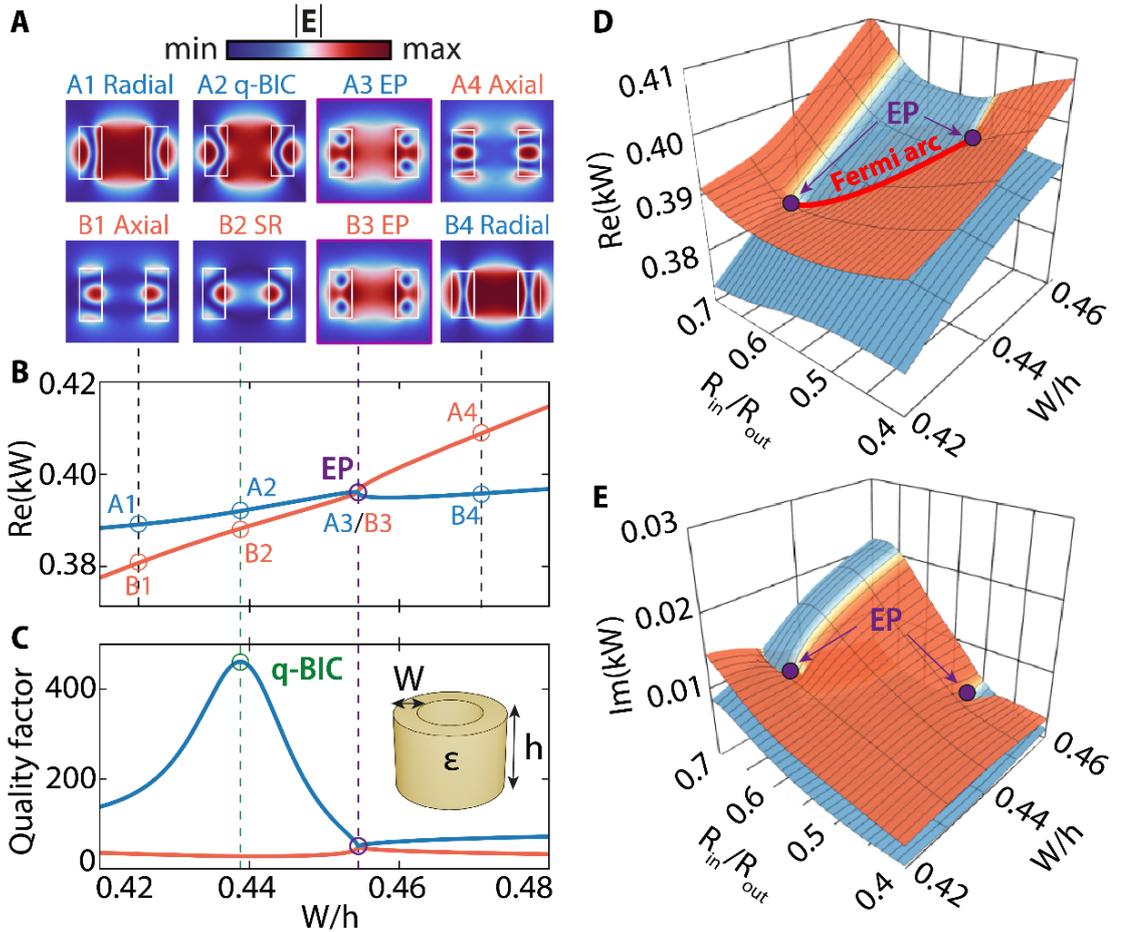

**Fig. 2 | Exceptional points and bulk Fermi arc in the single RR.** (A) Calculated near field distributions of $|E|$ in RR for both modes for the corresponding points in (B): at q-BIC, at EP, and at a larger spectral splitting $\Delta\omega$. (B) Real component of dimensionless eigenfrequency Re($kW$) relative to width-to-height ratio $W/h$. C Dependencies of quality factors of modes on the ratio $W/h$. (B) and (C) both constructed for a fixed radius ratio $R_{in}/R_{out} = 0.499$. (D)

and **(E)** Double Riemann sheet: surfaces of real and imaginary parts of eigenfrequencies kW in the parametric space. Purple dots mark the positions of EPs, and the red line in **(D)** marks the Fermi arc. The permittivity of the RR ε = 80.05.

**The "dark" nature of EPs: Quasinormal mode theory and Fano resonance**

EPs are mostly studied in terms of eigenvalues, as shown in Figures 1 and 2, but we also can find important peculiarities in individual spectral responses of eigenmodes near EP. For a single resonator with cylindrical symmetry illuminated by a linearly polarized plane wave, in which wave vector **k** is parallel to the symmetry axis, only photonic modes with an azimuthal index m = 1 can be excited. The extinction cross-section spectra, which are equal to the sum of the scattering and absorption cross-sections in photonics, are described by the Fano formula, the parameters of which can be analytically determined using the quasinormal mode (QNM) theory [34-36]. A single n[th] eigenmode (or QNM) produces a spectral response, which can be written in the next form:

$$\sigma_{ext}^n = \sigma_m \left[ \frac{q_n^2 - 1 + 2q_n\Omega_n}{(\Omega_n^2+1)(q_n^2+1)} \right], \quad (4)$$

where $\sigma_n$ is the intensity and $q_n$ is the Fano asymmetry parameter produced by n[th] mode; $\Omega_n = 2(\omega - \omega_n)/\gamma_n$ is dimensionless frequency, $\omega_n$ and $\gamma_n$ are the resonant frequency and half-width of the n[th] mode, respectively. The total extinction cross-section spectrum is the sum of QNM contributions (Eq. (4)) and background contribution $\sigma_{ext} = \sum_n \sigma_{ext}^n + \sigma_{bg}$ with the background term $\sigma_{bg} = \omega/(2P_0)Im\{\int_{V_r} \Delta\varepsilon \mathbf{E}_b^* \mathbf{E}_b d^3\mathbf{r}\}$. All parameters of a line shape (Eq. (4)) can be exactly calculated using the following formulas, according to [36]:

$$q_n = -\cot\left(\frac{\arg(\xi_n)}{2}\right), \quad (5)$$

$$\sigma_n = \frac{1}{P_0}\frac{\omega^2}{2\gamma_m}|\xi_n|, \quad (6)$$

$$\xi_n \equiv \left(\int_{V_{r'}} \Delta\varepsilon(\omega, \mathbf{r}')\tilde{\mathbf{E}}_n(\mathbf{r}') \cdot \mathbf{E}_b(\omega, \mathbf{r}')d^3\mathbf{r}'\right) \cdot \left(\int_{V_r} \Delta\varepsilon(\omega, \mathbf{r})\tilde{\mathbf{E}}_n(\mathbf{r}) \cdot \mathbf{E}_b^*(\omega, \mathbf{r})d^3\mathbf{r}\right), \quad (7)$$

where $\xi_n$ is the product of two independent overlap integrals; $\tilde{\mathbf{E}}_n$ is the normalized eigenmode field; $\mathbf{E}_b$ is the background field, which in our case corresponds to the incident plane wave field; $P_0$ is the time-averaged Poynting vector of the incident wave, and $\Delta\varepsilon = \varepsilon_0(\varepsilon_{res} - \varepsilon_b)$ is the dielectric contrast with resonator permittivity $\varepsilon_{res}$ and host media permittivity $\varepsilon_b = 1$; $\varepsilon_0$ is the vacuum permittivity.

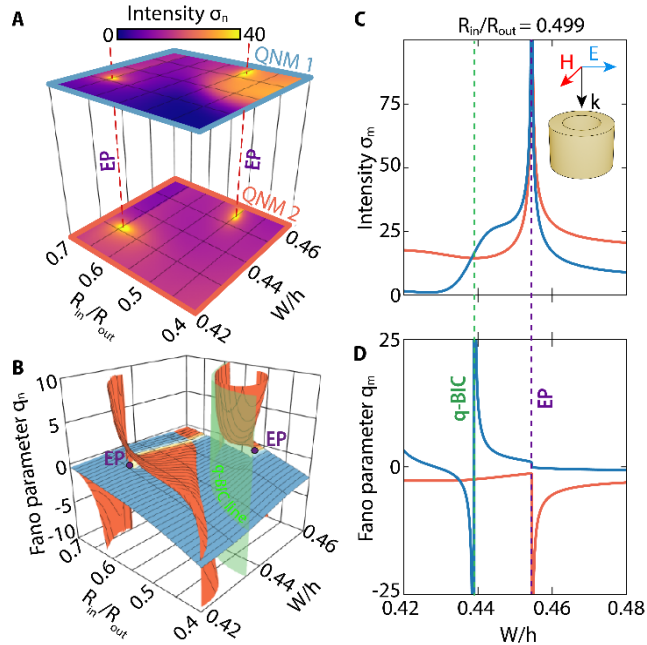

**Fig. 3 | Intensity and Fano parameter of EPs in parametric space at fixed excitation conditions. (A)** Behavior of intensities $\sigma_n$ for both eigenmodes near EPs in the parametric space ($R_{in}/R_{out}$, $W/h$), calculated using Eq. (6). **(B)** Behavior of the Fano parameters $q_n$ for both eigenmodes near EPs in the parametric space ($R_{in}/R_{out}$, $W/h$), calculated using Eq. (5). The green transparent surface corresponds to the q-BIC observation condition. **(C)** and **(D)** slices of surfaces **(A)** and **(B)** along $R_{in}/R_{out}$ = 0.499, respectively, at which the first EP is observed. Excitation conditions: linearly polarized wave with **k** parallel to the symmetry axis of the RR. The inset at C shows the incident wave.

Equations 5-7 allow us to study the behavior of the intensities $\sigma_n$ (Fig. 3A) and Fano parameters $q_n$ (Fig. 3B) for both modes independently in the parametric space ($R_{in}/R_{out}$, $W/h$). Figures 3A and 3C show that in the vicinity of EP, the intensities of both modes increase rapidly and become almost equal. This phenomenon is related to a near-zero normalization constant for modes that are close to coalescence, and the appearance of self-orthogonality is thus another well-known marker of an EP [37] (see Supplementary Information, section II). Similar results for the Fano asymmetry parameter $q_n$ was obtained using Eq. (5). The blue surface in Figure 3B corresponds to one of the QNMs (blue lines in Figs. 3C and 3D), and it has multiple divergences, where the resonant response becomes Lorentz-like ($q_n \to \pm\infty$). Such an infinite value of the Fano parameter, in some cases, marks the positions of the q-BIC regime, the observation condition of which is demonstrated in Figure 3B by the green transparent surface. For other combinations of parameters ($R_{in}/R_{out}$, $W/h$), the Fano line has an asymmetric shape with $|q_n| <$ 15. Meanwhile, the second QNM corresponding to the orange surface in Figure 3B remains asymmetric almost everywhere in the chosen parametric range with the Fano parameter $q_n \approx -1$, except for the vicinity of the EP (Figs. 3B and 3D). An important result is that the Fano parameters $q_m$ of two interacting modes at the EP can be: 1) equal to one in absolute value but have opposite signs; 2) equal to zero for one mode and $\pm\infty$ for the other (Fig. 3D). These conditions correspond to total destructive interference and explain the "dark" nature of the EP, which does not manifest itself in the extinction spectra in any way despite the very high intensity of both modes that form the EP.

## Experimental observation of the bulk Fermi arc

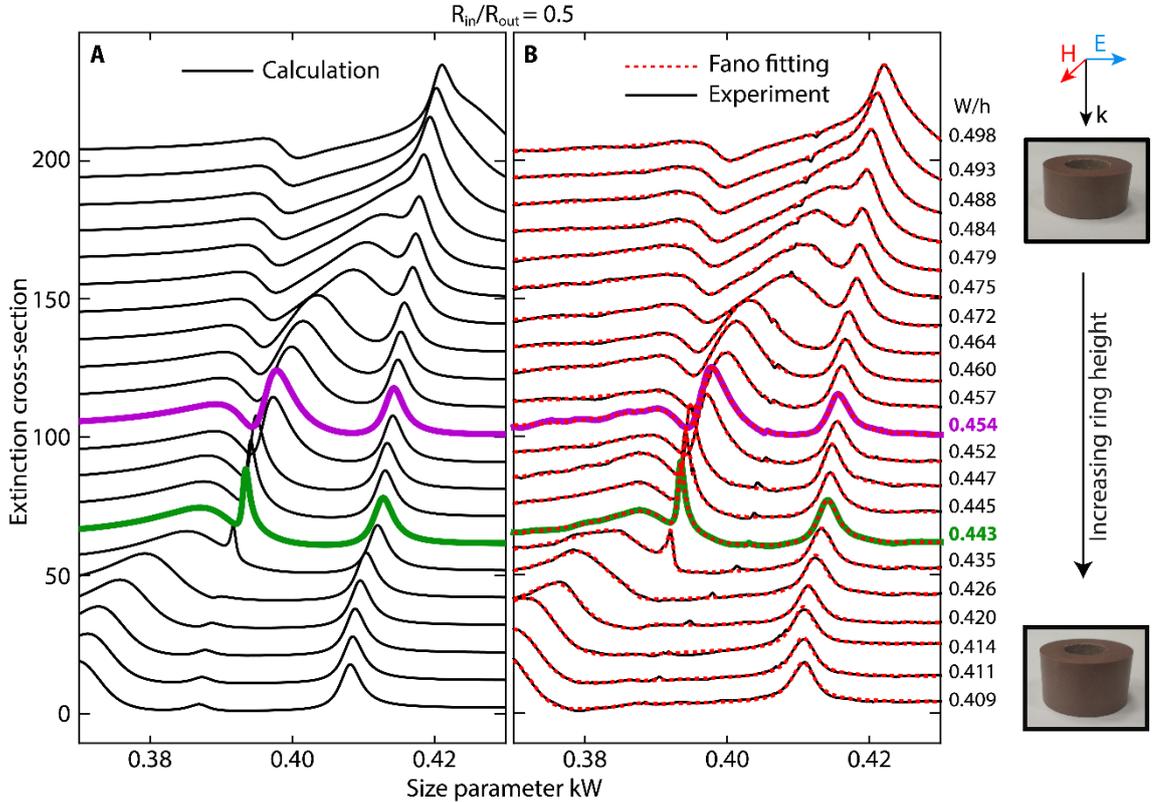

**Fig. 4 | Experimental extinction spectra of RRs, results of calculations and fitting. (A)** Calculated extinction cross-section spectra for fixed $R_{in}/R_{out}$ = 0.5 with variable ratio $W/h$. **(B)** Experimentally measured (black lines) and

fitted (red dashed lines) extinction cross-section spectra for $R_{in}/R_{out}$ = 0.5. The spectrum corresponding to the q-BIC ($W/h$ ~ 0.439) is highlighted with green color. The spectrum corresponding to the EP ($W/h$ ~ 0.455) is highlighted with purple color. The excitation setup and the scheme of changing the RR height h are shown on the right panel. The permittivity of RRs $\varepsilon$ = 80.05.

Comprehensive experimental studies of EP in dielectric RR are presented below and include experimental observations of Riemann sheets, Fermi arc, and topological features of EP. For the studies, we used thirteen RR samples to obtain the experimental eigenfrequencies in the parametric space. The samples had a fixed $R_{in}/R_{out}$ and the highest height $h$ corresponding to $W/h$ = 0.420 (Methods, Fig. 6). The height $h$ of each experimental sample was gradually reduced to $W/h$ = 0.46. Depending on the value of the ring width $W$, it is necessary to have a different initial height $h$ and height step $\Delta h$. Thus, for each of the samples, a different number of experimental spectra were made (Supplementary Information, section III), resulting in 203 measurements in total. As an example, Figure 4B demonstrates a series of experimentally measured extinction spectra for samples with variable height and a fixed ratio $R_{in}/R_{out}$ = 0.5, which is the closest parameter to EP obtained in numerical calculations for $R_{in}/R_{out}$ ≈ 0.499. The experimental spectra (Fig. 4B) are in excellent agreement with the calculated ones (Fig. 4A), and this agreement was observed for all 13 series of experiments for different $R_{in}/R_{out}$ ratios (all spectra are presented in the Supplementary Information, section III). The accurate fit of experimental spectra is provided with theoretically precalculated parameters – eigenfrequencies, intensities, and Fano parameters, that were used for the initial steps of the Fano fitting procedure (see Supplementary Information, section III). Otherwise, fitting near the EP is extremely difficult without an accurate theory predicting the correct function that should be used for the fitting. The red dashed lines in Figure 4B correspond to the Fano fitting of the experimental spectra, which are in excellent agreement with the experimental curves.

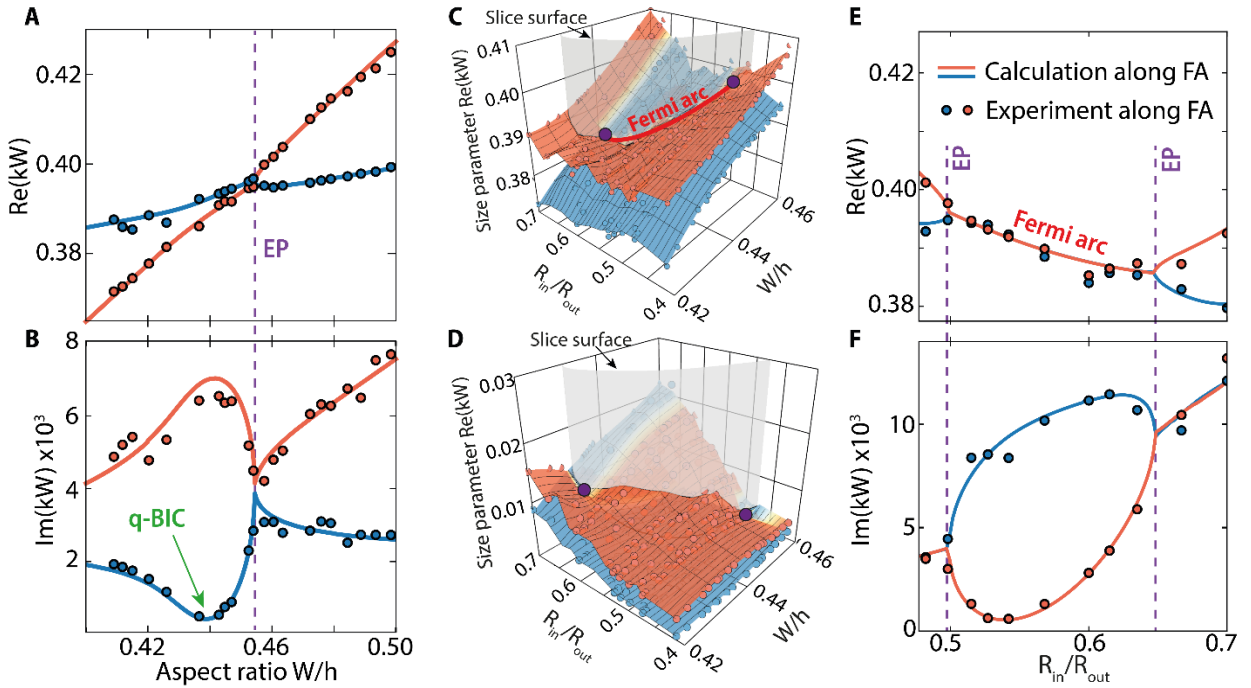

**Fig. 5 | Experimental observation of the Fermi arc in dielectric RRs.** (**A**, **B**) Dependences of the real and imaginary parts of the dimensionless frequency $kW$ of two RRs modes on the aspect ratio $W/h$, respectively. The data obtained by fitting the experimental spectra are represented as dots. RR has a fixed radius ratio $R_{in}/R_{out}$ = 0.5. (**C**, **D**) Double Riemann sheet for the real and imaginary parts of the eigenvalues, respectively, obtained by connecting the experimental points (shown as colored dots). The two purple dots show the approximate positions of the experimental EPs. The red line shows the resulting experimental bulk Fermi arc. (**E**) Experimental observation of the bulk Fermi arc and (**F**) corresponding dependence for the imaginary part of eigenfrequency as a function of $R_{in}/R_{out}$. The permittivity of RRs $\varepsilon$=80.05.

The results of experimental data fitting (resonant frequencies and half-widths) were then compared with calculated ones. For RR with $R_{in}/R_{out} = 0.5$, Figures 5A and 5B demonstrate the real and imaginary parts of two dimensionless eigenfrequencies $kW$ on the $W/h$ ratio. The height of the sample was changed 21 times; fitted parameters are depicted as colored dots in Figs. 5A and 5B. We observed a good agreement between the calculated (solid lines, QNM eigenfrequencies) and experimental results (dots). The EP corresponds to the value of $W/h \approx 0.454$, and the q-BIC occurs at $W/h \approx 0.439$. A sensing performance of two EP points depending on the perturbation is carried out in [33]. Fitting data for all RRs is given in the Supplementary Information, section III.

For experimental observation of the bulk Fermi arc, a height reduction procedure was applied to all thirteen samples; each extinction spectrum was fitted as described above (see Supplementary Information, section III). Figures 5C and 5D demonstrate the main result of the work – experimental observation of the bulk Fermi arc. The experimental double Riemann sheets in the parametric space ($R_{in}/R_{out}$, $W/h$) were obtained by combining the fitting data of all experimental spectra. The red line in 5C marks the position of the bulk Fermi arc in the dielectric RRs obtained on the basis of experimental data. This bulk Fermi arc was obtained by interpolating the intersection points of the sheets of real or imaginary parts of the eigenvalues of the numerically calculated eigenfrequencies. The results of this process are shown in Figures 5C and 5D as the values of the "cut surface". A comparison of the eigenvalues obtained from experiments and calculations is shown in Figures 5E and 5F. In region $0.5<R_{in}/R_{out}<0.65$, both numerical and experimental data perfectly coincide and show the presence of a bulk Fermi arc.

In order to show the topological nature of EP, we consider closed curves around EP in the parametric space for both calculated (black curve) and experimental (light green curve) results (Fig. 6A). Upon circulation along these curves, the eigenstates exchange the Riemann sheets when crossing the Fermi arc. Therefore, to return to the initial position, it is necessary to make not a trivial $2\pi$-trip but a $4\pi$-trip around the EP (Fig. 6B). The experimental eigenvalues in Fig. 6D are plotted for the corresponding points, numbered in Fig. 6A in the polygonal light green curve since a circular trajectory is difficult to determine from experiments. Both the calculated (Fig. 6C) and experimental (Fig. 6D) spectra for the numbered points are in good agreement. The behavior demonstrated in Fig. 6B corresponds to the classical trajectory on the surface of the 1D Mobius strip, for which a $4\pi$ round trip is necessary to return to the initial point. This feature is a unique marker of the topological nature of EPs [2, 4, 6].

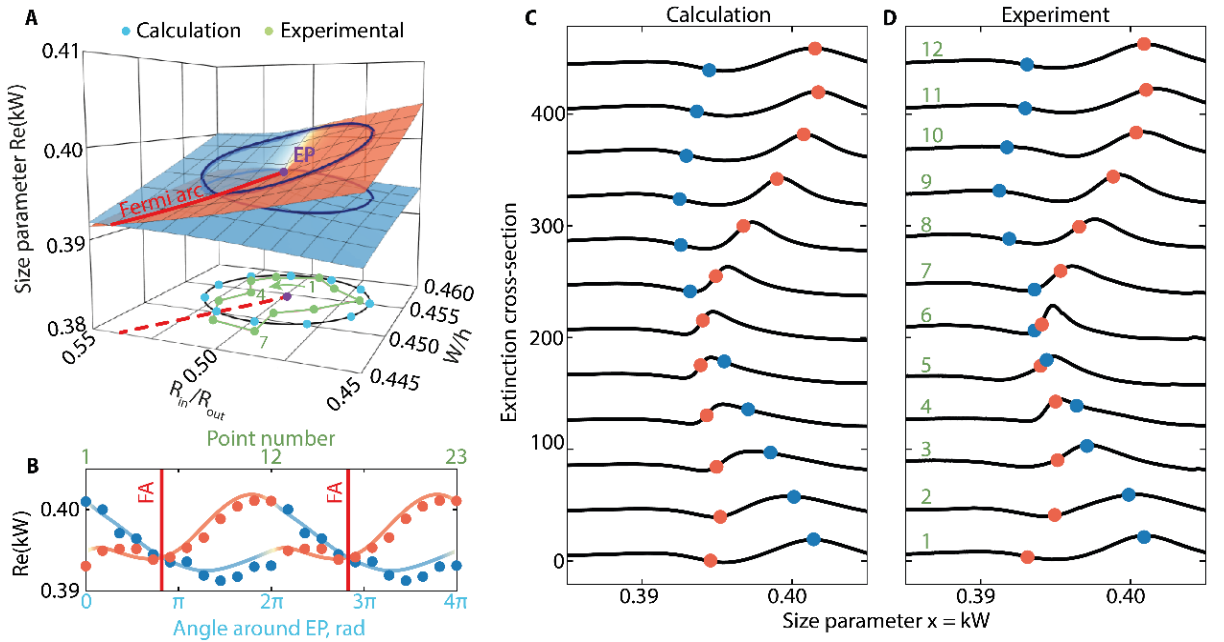

**Fig. 6 | Transformation of eigenvalues and spectra upon a circular trip around EP. (A)** The calculated surface of the real part of the eigenvalues including EP at $R_{in}/R_{out}$ = 0.499, $W/h$ = 0.455 (marked with a purple dot). A dark blue line on the surface shows the behavior of the eigenvalues around EP. The geometrical parameters of the ring, for which the spectra are plotted, are shown by cyan dots in the ($R_{in}/R_{out}$, $W/h$) plane. The dots form a closed curve around EP. The experimental light green dots are numbered. The red dotted line represents the projection of the Fermi arc onto the plane **(B)** The $4\pi$ round trip of real part of eigenvalues around EP and the corresponding experimental values plotted as a function of the point number and corresponding polar angle in the horizontal plane (A). Note that the angle starts counting from the point №1 in counterclockwise direction. A double pass is demonstrated. **(C, D)** Calculated and experimental extinction spectra and the corresponding eigenvalues marked with colored dots. The permittivity of RRs $\varepsilon$ = 80.05.

## Discussion

In non-Hermitian physics, the emergence of exceptional points is one of the most striking and universal characteristics [11, 38]. Here, we demonstrated the experimental double Riemann sheet and Fermi Arc for a single dielectric RR with a rectangular cross-section. A pair of EPs connected by a bulk Fermi arc that formed a topologically stable "dumbbell" configuration [4] was presented. Experimental studies of the eigenfrequencies in the far-field extinction spectra were carried out on 203 ceramic RRs (Methods, Experimental studies). Experimental results were supported by comprehensive theoretical studies based on QNM theory. We showed that the EPs are "dark" states (Fig. 3C) due to the coincidence of the amplitudes and line shapes of the two modes forming the EPs with opposite phases. The proximity of EPs and q-BIC in the parametric space was also demonstrated experimentally (Fig. 4), which corresponds to previously observed results [27].

Our experimental results on the observation of a bulk Fermi arc in a single compact resonator, as well as the methodology of the theoretical study, provide a platform for the future exploration of non-Hermitian and topological physics in structures of different dimensions and complexity and point to non-trivial ways of controlling and engineering light-matter interaction [24, 39].

## Methods

### Theoretical calculations

The numerical simulations were performed using COMSOL Multiphysics. For a detailed physical analysis, we applied the QNM theory [40, 41]. QNMs are source-free solutions of Maxwell's equations that were obtained using COMSOL eigensolver and normalized using the PML norm [42]. The QNM theory allows for the calculation of all spectral response parameters, including the scattering intensities $\sigma_m$ and the Fano parameters $q_m$ [36], thus obtaining 3D $\sigma_m$ and $q_m$ maps in the parametric space. Each 3D map consists of 301 rows along $R_{in}/R_{out}$ in the range from 0.4 to 0.7 and 601 columns along $(R_{out}-R_{in})/h = W/h$ in the range from 0.42 to 0.46. The numerical extinction cross-section spectra $\sigma_{ext}$ were obtained as a linear sum of the responses of single QNM responses under plane wave excitation (50 QNMs were used). The extinction cross-section $\sigma_{ext}$ was normalized to the geometric shadow, which for the dielectric RR is equal to $S = 2R_{out}h$.

### Experimental studies

The experiments were performed on BaO−Ln$_2$O$_3$−TiO$_3$ ceramic samples with a permittivity of $\varepsilon \approx 80.05$ and a loss tangent of $\tan(\delta) = 1.8 \cdot 10^{-4}$. The material used lends itself well to mechanical changes in geometry. The samples are centimeter-sized, each with an outer radius of $R_{out}$ = 16 mm, which allows changing the $W/h$ parameter with high accuracy. First, RRs with specified $R_{in}/R_{out}$ ratios (Fig. 7) and a height corresponding to $W/h$ = 0.42 were made from cylinders, after that the height of each RR was smoothly decreased. In total, 13 samples with different $R_{in}/R_{out}$ ratios in the range from 0.4 to 0.7 were used. The RRs were changed in height an average of 15 times, which yielded a total of 203 RRs and, accordingly, 203 experimental spectra in the parametric space ($R_{in}/R_{out}$, $W/h$).

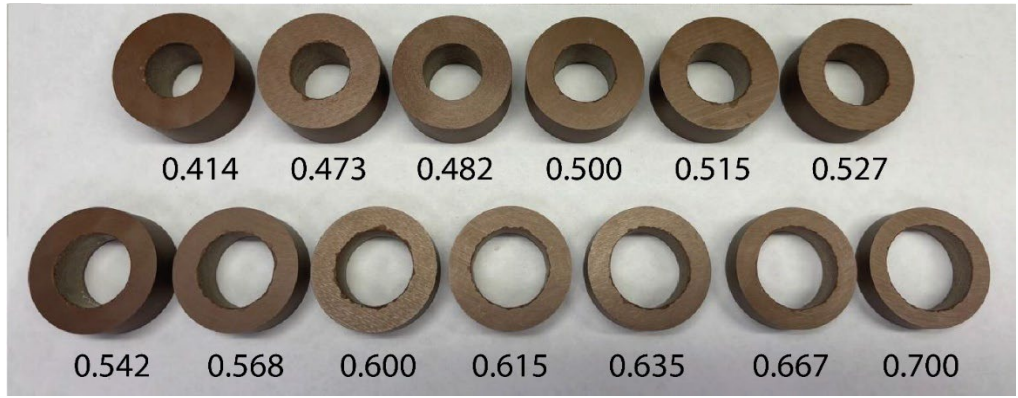

**Fig. 7** | Photograph of 13 RRs forming a "two-dimensional matrix" of 203 RRs used in the experiments. Changing the height of each RR allows for covering the entire parametric space ($R_{in}/R_{out}$, $W/h$) containing the bulk Fermi arc. The ratio of radii $R_{in}/R_{out}$ is indicated under each RR.

Far-field experiments were carried out in a microwave anechoic chamber in the range of 1-6 GHz [31, 43]. For the far-field extinction measurements, the sample was placed on a custom mount, transparent in the microwave range, between two horn antennas at a distance of ~2 meters from each other. One antenna generated a plane wave with a **k** vector along the axial axis of RR, and the other received the transmitted signal. The transmitting and receiving horns were connected to a vector network analyzer Rohde & Schwarz ZVB20. This setup represents a two-port system that can be described by scattering matrix $S$.

To measure the extinction cross-section, it is sufficient to obtain forward signals twice, with and without a sample. The extinction spectrum can be obtained using the optical theorem [44]:

$$\sigma_{ext} = -\frac{4\pi c}{\omega}\frac{L_r L_l}{L} Im\left(\frac{S_{12}}{S_{12}^b}\right), \tag{8}$$

where $c$ is the speed of light, $\omega$ is the angular frequency, $L$ is the distance between antennas, $S_{12}$ is the measured transmission spectrum from the samples, $S_{12}^b$ and is the measured transmission spectrum in the absence of the sample.

## Acknowledgments

The authors thank Pavel Belov and Yuri Kivshar for the discussion and Elizaveta Nenasheva (Ceramics Co. Ltd., St. Petersburg) for providing ceramic samples for measurements. The theoretical studies were supported by the Russian Science Foundation (Project 23-72-10059). The experimental studies were supported by the Russian Science Foundation (Project 23-12-00114). N.S. acknowledges support from the Advancement of Theoretical Physics and Mathematics "BASIS" Foundation.

## Author contributions
N.S. and F.Z. performed all experiments. N.S. performed all calculations and experimental data processing. M.B and N.S. studied the responses of quasinormal modes near singular points (q-BIC and EP). K.B. and M.S. described the mode coupling using temporal coupling mode theory. M.L., A.B., M.B., and N.S. analyzed the theoretical and experimental results and co-wrote the draft of the paper. M.L served as the project supervisor. All authors discussed and contributed to the project.

## Competing interests
The authors declare no competing interests.

## Additional information
Supplementary information The online version contains supplementary material available at https://doi.org.......
Correspondence and requests for materials should be addressed to A. Bogdanov.

# Supplementary Information for

# Experimental observation of bulk Fermi arc in single dielectric resonator

N. Solodovchenko,[1,2,3] F. Zhang,[1,2] M. Bochkarev,[2] K. Samusev,[2,3] M. Song[1], A. Bogdanov,[1,2] and M. Limonov[2,3]

[1]*Qingdao Innovation and Development Center, Harbin Engineering University, Qingdao, 266000, Shandong, China.*
[2]*School of Physics and Engineering, ITMO University, Saint-Petersburg 197101, Russia*
[3]*Ioffe Institute, St. Petersburg, 194021, Russia*

**Content:**

**I. Derivation of the conditions for appearing of the q-BIC and EP**

**II. Bi-orthogonality, normalization and self-orthogonality of eigenmodes in the EP state**

**III. Fitting of spectra**

**I. Derivation of the conditions for appearing of the q-BIC and EP**

In this section we derive the necessary conditions for the appearance of the q-BIC and EP from the two-level Hamiltonian for two resonances (or eigenmodes), when they are coupled to the same radiation channel (Fig. S1).

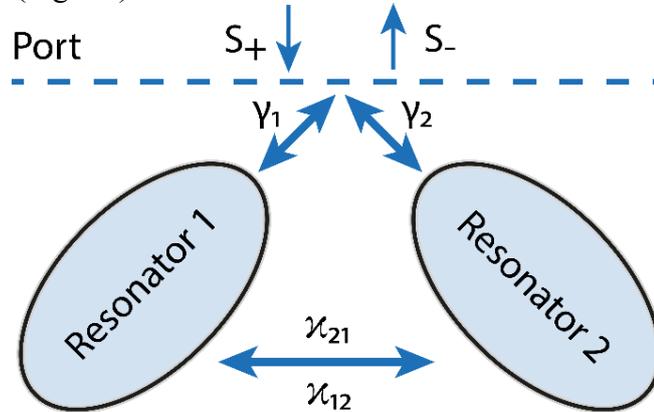

**Fig. S1** – A system of two resonators (or two eigenmodes) that couples to the same radiation channel.

Such a system can be described by the following Hamiltonian:

$$H = \begin{pmatrix} \omega_1 & \varkappa \\ \varkappa & \omega_2 \end{pmatrix} - i \begin{pmatrix} \gamma_1 & \sqrt{\gamma_1 \gamma_2} \\ \sqrt{\gamma_1 \gamma_2} & \gamma_2 \end{pmatrix}, \qquad (S1)$$

and characteristic equation for this Hamiltonian:

$$\widetilde{\omega}^2 - [(\omega_1 + \omega_2) + i(\gamma_1 + \gamma_2)]\widetilde{\omega} + [(\omega_1 \omega_2 - \varkappa^2) + i(\gamma_1 \omega_2 + \gamma_2 \omega_1 - 2\varkappa\sqrt{\gamma_1\gamma_2})] = 0, \qquad (S2)$$

For that moment all notations corresponded to those from the main text, now we introduce new notations for simplicity: $\omega_0 = (\omega_1 + \omega_2)/2$; $\Omega = \omega_1 - \omega_2$; $\gamma_s = \gamma_1 + \gamma_2$; $\gamma_d = \gamma_1 - \gamma_2$; $\gamma_0 = \sqrt{\gamma_1 \gamma_2}$. Then, Eq. (2) can be rewritten in the next form:

$$\left[\widetilde{\omega} - \left(\omega_0 + i\tfrac{\gamma_s}{2}\right)\right]^2 = \tfrac{1}{4}[(\Omega^2 + 4\varkappa^2 - \gamma_s^2) + 2i(\Omega\gamma_d + 4\varkappa\gamma_0)], \tag{S3}$$

with the solution for eigenfrequency:

$$\widetilde{\omega} = \left(\omega_0 + i\tfrac{\gamma_s}{2}\right) + \tfrac{1}{2}\sqrt{(\Omega^2 + 4\varkappa^2 - \gamma_s^2) + 2i(\Omega\gamma_d + 4\varkappa\gamma_0)}. \tag{S4}$$

Let's consider this root in detail. Here we introduce by definition two quantities: $\equiv \Omega^2 + 4\varkappa^2 - \gamma_s^2$; $b \equiv 2(\Omega\gamma_d + 4\varkappa\gamma_0)$. From the basic complex analysis, it comes that: $r + is = \sqrt{a + ib} \Rightarrow (r + is)^2 = a + ib \Rightarrow a = r^2 - s^2$; $b = 2rs$. Next, we substitute $s = b/(2r)$ into equation for $a$ and obtain biquadratic equation for $r$:

$$r^2 - \tfrac{b^2}{4r^2} - a = 0 \Rightarrow r^4 - ar^2 - \tfrac{b^2}{4} = 0, \tag{S5}$$

with the next solutions for $r$ and $s$:

$$r_\pm = \pm\tfrac{1}{\sqrt{2}}\sqrt{a + \sqrt{a^2 + b^2}} \;;\; s_\pm = \tfrac{b}{2r_\pm}. \tag{S6}$$

Note that $r$ is a real number and $r^2$ cannot be negative, due to this we take just one of solutions of Eq. (S5) and use it to obtain two solutions in Eq. (S6). As a result, we obtain two eigenfrequencies:

$$\widetilde{\omega}_\pm = \left(\omega_0 \pm \tfrac{1}{2\sqrt{2}}\sqrt{a + \sqrt{a^2 + b^2}}\right) + i\tfrac{1}{2}\left(\gamma_s \pm \tfrac{\sqrt{2}b}{\sqrt{a + \sqrt{a^2 + b^2}}}\right). \tag{S7}$$

Now we will study behavior of the imaginary part $\tfrac{b}{\sqrt{a + \sqrt{a^2 + b^2}}}$ with respect to $\Omega$ as a variable. We can find the roots for the derivative $\partial\left(\tfrac{b}{\sqrt{a + \sqrt{a^2 + b^2}}}\right)/\partial\Omega = 0$, and obtain two corresponding solutions: $\Omega_1 = \tfrac{\varkappa\gamma_d}{\gamma_0}$ and $\Omega_2 = -\tfrac{\gamma_0\gamma_d}{\varkappa}$. Substitution of the $\Omega_1$ in the derivative will provide an exact zero, and the $\widetilde{\omega}$ at this condition always have an extremum value in the imaginary part. Therefore, substitution of $\Omega_1$ in Eq. (S7) for $\omega_\pm$ correspond the q-BIC with coupling coefficient:

$$\varkappa_{q-BIC} = \tfrac{(\omega_1 - \omega_2)\sqrt{\gamma_1\gamma_2}}{(\gamma_1 - \gamma_2)}. \tag{S8}$$

From the other side, we can study behavior of the real part $\sqrt{a + \sqrt{a^2 + b^2}}$ with respect to $\Omega$. We can find roots for derivative $\partial\left(\sqrt{a + \sqrt{a^2 + b^2}}\right)/\partial\Omega = 0$ and discover that $\Omega_2 = -\tfrac{\gamma_0\gamma_d}{\varkappa}$ correspond to minimum value in real part of eigenfrequency, which corresponds to the Rabi point with: $\Omega_{Rabi} = Re\{\widetilde{\omega}_+\} - Re\{\widetilde{\omega}_-\} = \sqrt{4\varkappa^2 - \gamma_d^2}$.

From Eq. (S7) it is straightforward that conditions for the EP, correspond to $\sqrt{a + \sqrt{a^2 + b^2}} = 0$ and $\sqrt{2}b/\sqrt{a + \sqrt{a^2 + b^2}} = 0$, simultaneously; at the same time it is necessary to $a = 0$ and $b = 0$. Therefore, conditions for the EP:

$$\begin{cases} b \equiv 2(\Omega\gamma_d + 4\varkappa\gamma_0) = 0 \Rightarrow \varkappa_{EP} = -\tfrac{(\omega_1 - \omega_2)(\gamma_1 - \gamma_2)}{4\sqrt{\gamma_1\gamma_2}}, \\ \Omega_{Rabi} = \sqrt{4\varkappa^2 - \gamma_d^2} = 0 \Rightarrow \varkappa_{EP} = \pm|\gamma_1 - \gamma_2|/2. \end{cases} \tag{S9}$$

When $|\varkappa| < |\gamma_1 - \gamma_2|/2$ the weak coupling of modes and the intersection are observed (only in real or imaginary part), and $|\varkappa| > |\gamma_1 - \gamma_2|/2$ represents the strong coupling and anticrossing. In the case of interaction of two modes in a single dielectric resonator, EPs always appear in pairs, which corresponds to different signs of $\varkappa_{EP}$. These two EPs are connected in parametric space by the bulk Fermi arc, along which real parts of eigenfrequencies are equal.

For the demonstration, we plotted the results for $\widetilde{\omega}_\pm$ for fixed $\gamma_1 = 0.7$ and $\gamma_2 = 0.2$ (Fig. S2). The parameter $\alpha$ is related to frequencies of the uncoupled resonators in the next way: $\omega_1 = \alpha$, $\omega_2 = 3\alpha$. We can consider $\alpha$ for example, as some geometric parameter of the resonator, the changes of which have different impact on uncoupled frequencies ($\omega_1$ and $\omega_2$).

Such different impact can be observed for radial and axial modes in a RR upon changes in height – for radial modes impact on eigenfrequency is less than for axial (as demonstrated in the main text), thus such relation between α and ($\omega_1$, $\omega_2$) is a realistic simplification. Fig. S2(a) shows behaviors of eigenfrequencies $\widetilde{\omega}_\pm$ and the bulk Fermi Arc at which $|\varkappa| < |\gamma_1 - \gamma_2|/2$ (blue line) and two EPs, where $\varkappa_{EP} = \pm|\gamma_1 - \gamma_2|/2$. Fig. S2(b, c) shows real and imaginary parts of eigenfrequencies $\widetilde{\omega}_\pm$ at $\varkappa = -0.25$, which corresponds to one of the EPs. When $Im\{\widetilde{\omega}_+\} \to 0$, we observe q-BIC, which occurs at the condition from Eq (S8).

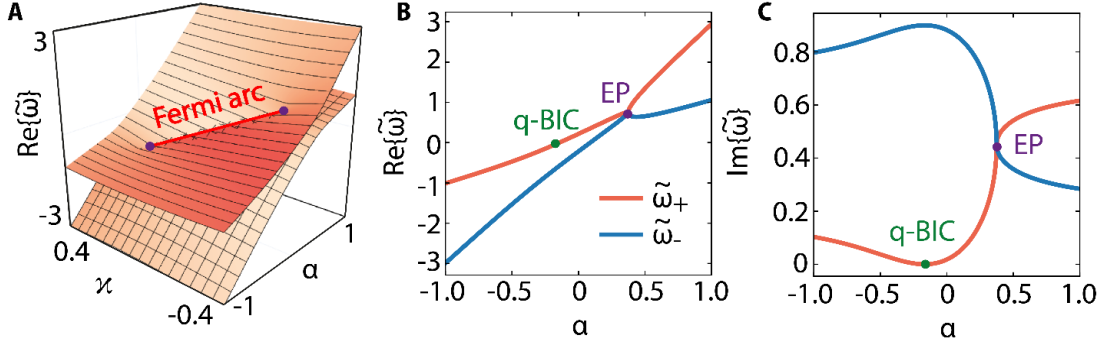

**Fig. S2** – Solutions for Hamiltonian (Eq. (S1)). (**A**) Eigenfrequencies $\widetilde{\omega}_\pm$ in parametric space and the Fermi arc (red line). (**B**) and (**D**) Spectral positions of real and imaginary parts of eigenfrequencies $\widetilde{\omega}_\pm$ with marked EP and q-BIC.

Equation S9 shows that if $|\gamma_1 - \gamma_2|/2$ decreases, the Fermi arc becomes smaller and in the limiting case, when the losses of the two resonances are equal, degenerates into a Dirac point (DP). Figure S3 shows several examples for different values of resonance losses.

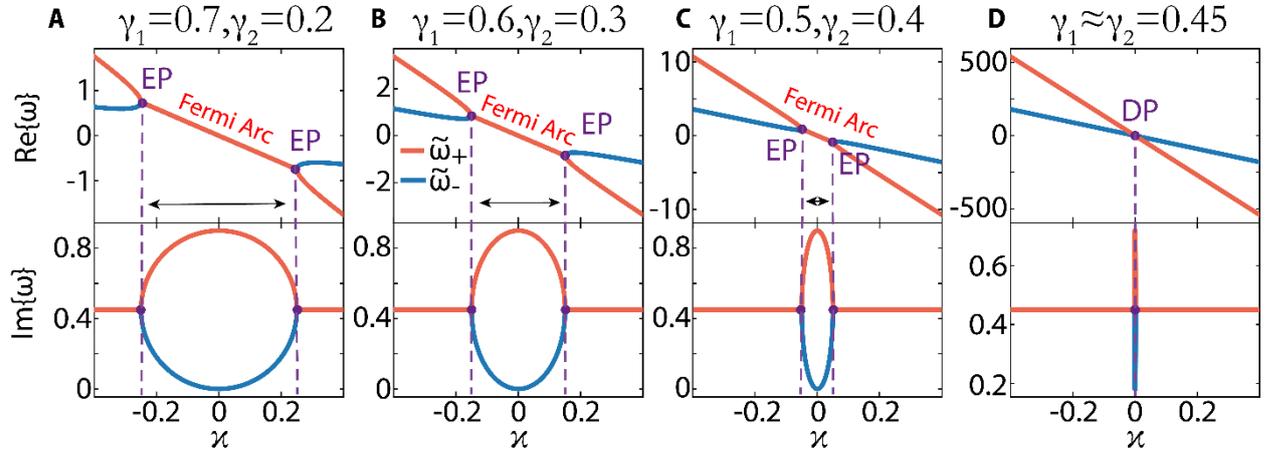

**Fig. S3** – Solutions for Hamiltonian (Eq. (S1)) along Fermi arc with different values of resonance losses. (**A**) Real and imaginary parts of eigenfrequencies $\widetilde{\omega}_\pm$ with different values of resonance losses (A) $\gamma_1 = 0.7$, $\gamma_2 = 0.2$, (**B**) $\gamma_1 = 0.6$, $\gamma_2 = 0.3$, (**C**) $\gamma_1 = 0.5$, $\gamma_2 = 0.4$ and (**D**) $\gamma_1 \approx \gamma_2 = 0.45$ marked above the figures.

## II. Bi-orthogonality, normalization and self-orthogonality of eigenmodes in the EP state

Two eigenmodes are source-free solutions of the Maxwell equations $\widehat{H}\widetilde{\Psi}_{v,n} = \widetilde{\omega}_{v,n}\widetilde{\Psi}_{v,n}$, with $\widetilde{\Psi}_n = (\widetilde{E}_n, \widetilde{H}_n)$. Biorthogonality relation is based on the non-conjugate form of the Lorentz reciprocity theorem [1]:

$$(\widetilde{\omega}_n - \widetilde{\omega}_v) \iiint_V \widetilde{\Psi}_n^T \widehat{D} \widetilde{\Psi}_v^T d^3 r = i \iint_\Sigma (\widetilde{E}_n \times \widetilde{H}_v - \widetilde{E}_v \times \widetilde{H}_n) \cdot d\mathbf{s}, \qquad (S10)$$

where $V$ is the integral over the entire volume including PML, $\Sigma$ is the integral over the PML surface, which is zero under PEC boundary conditions, and for the non-dispersive case $\widehat{D} =$ $diag(-\mu_0, \varepsilon_0, \varepsilon_{res})$, thus the right side of the equation will be zero. Thus, for $m \neq n$, the volume integral $\iiint_V \widetilde{\Psi}_n^T \widehat{D} \widetilde{\Psi}_v^T d^3r = 0$, while for $m = n$ condition become $\iiint_V \widetilde{\Psi}_n^T \widehat{D} \widetilde{\Psi}_v^T d^3r = 1$. The most general form of the PML norm for both dispersive and non-dispersive media can be obtained with equation (S11):

$$QN = \iiint_V \left[ E_n \cdot \frac{\partial \omega \varepsilon}{\partial \omega} E_n - H_n \frac{\partial \omega \mu}{\partial \omega} H_n \right] d^3r. \tag{S11}$$

The value of this integral for the non-normalized eigenmodes $(E_n, H_n)$ calculated in COMSOL by varying in the geometry parameter W/h with fixed value of $R_{in}/R_{out} = 0.499$ is shown in Figure S4. When approaching the EP, the integral value tends modulo zero, which leads to a large value of the normalized field $\widetilde{E}_n = E_n/\sqrt{QN}$. Substituting $\widetilde{E}_m$ in the equation (5) of the main text will result in a large value of $|\xi_n|$, which corresponds to a giant intensity of single QNM in the vicinity of the EP. The tendency of the integral QN to zero is explained by the manifestation of the self-orthogonality of two close eigenvalues both in terms of eigenvalue and the field distribution.

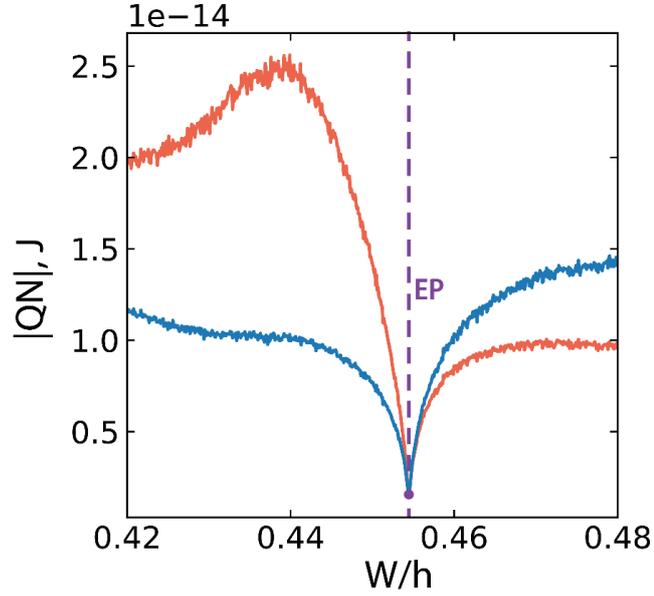

**Fig. S4** –PML-norm behavior near the EP.

The intensity of a single QNM is calculated along the Fermi arc, which was obtained by interpolating the intersection points of two QNM by a polynomial of the real or imaginary surface of the eigenvalues. Despite the fact that the intensities of the two QNM have large values, their total intensity, as a result of destructive interference, will remain at the same level.

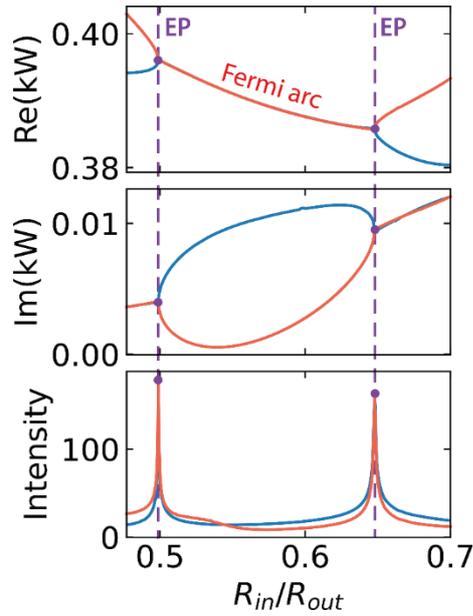

**Fig. S5** – The results of calculating the eigenvalues along the Fermi arc: the real, imaginary part of the eigenvalues and the corresponding extinction intensity of two QNM at normal incidence on the dielectric RR.

### III. Fitting of spectra

To fit the experimental spectra, the eigenvalues, intensity, and Fano parameters for a given incident wave for each experimental spectrum were first calculated for each experimental spectrum. The calculation showed that there are 3 eigenmodes in the frequency range of interest, which corresponds to 12 unknown parameters. Thus, for each experimental spectrum, 12 numerically calculated parameters were used as an initial approximation to find 12 experimental parameters. Figure S6 shows the experimental spectrum $R_{in}/R_{out} = 0.5$, $W/h = 0.494$, which was fitted (red dotted line) using parameters of 3 QNM (orange, green and red solid line), the sum of which corresponds to the experimental spectrum.

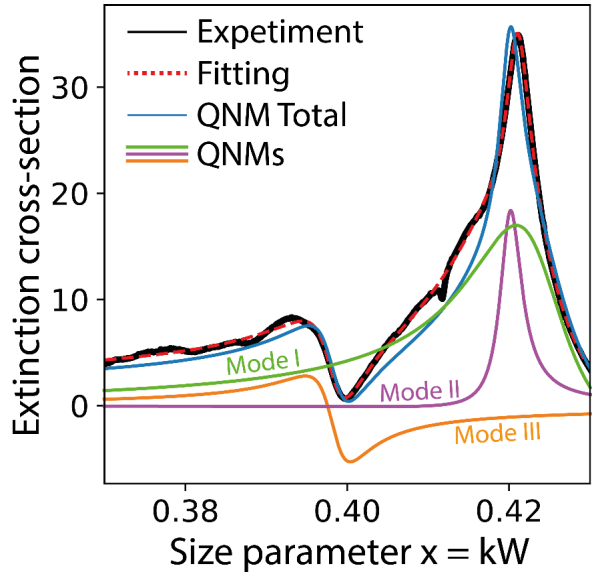

**Fig. S6** – Fitting of the experimental spectrum (black solid line) using 3 QNM parameters (orange, green and red solid line) as initial parameters $p_0 = 3 \times (\omega_m, \gamma_m, \sigma_m, q_m)$.

Such a fitting operation was applied to each of the extinction spectra obtained in the experiment. Examples of spectra fitting for each of the ring resonators are presented below.
Figures S7-S9 demonstrate:
- Top row: photographs of samples and parameter $R_{in}/R_{out}$.
- Second row: experimentally measured (black lines) and fitted (red dashed lines) extinction cross-section spectra. The spectrum corresponding to the q-BIC is highlighted in green, the one corresponding to the EP is highlighted in purple.
- Third row and fourth row - dependences of the real and imaginary parts of the dimensionless frequency of two RRs modes on the aspect ratio W/h, respectively.

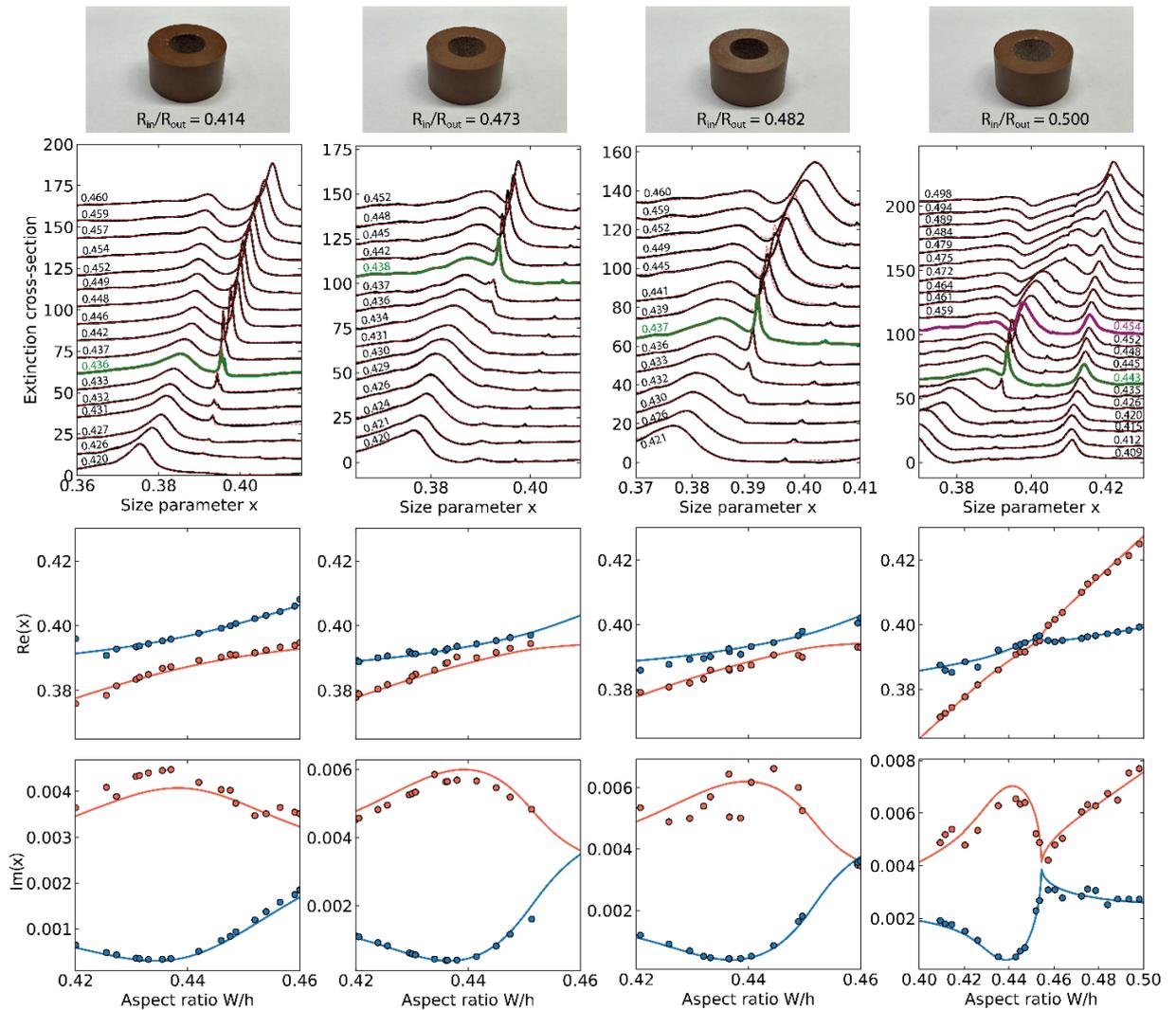

**Fig. S7** – Fitting of the experimental spectrum (black solid line) and eigenfrequencies vs W/h. Range of $0.4 < R_{in}/R_{out} < 0.5$.

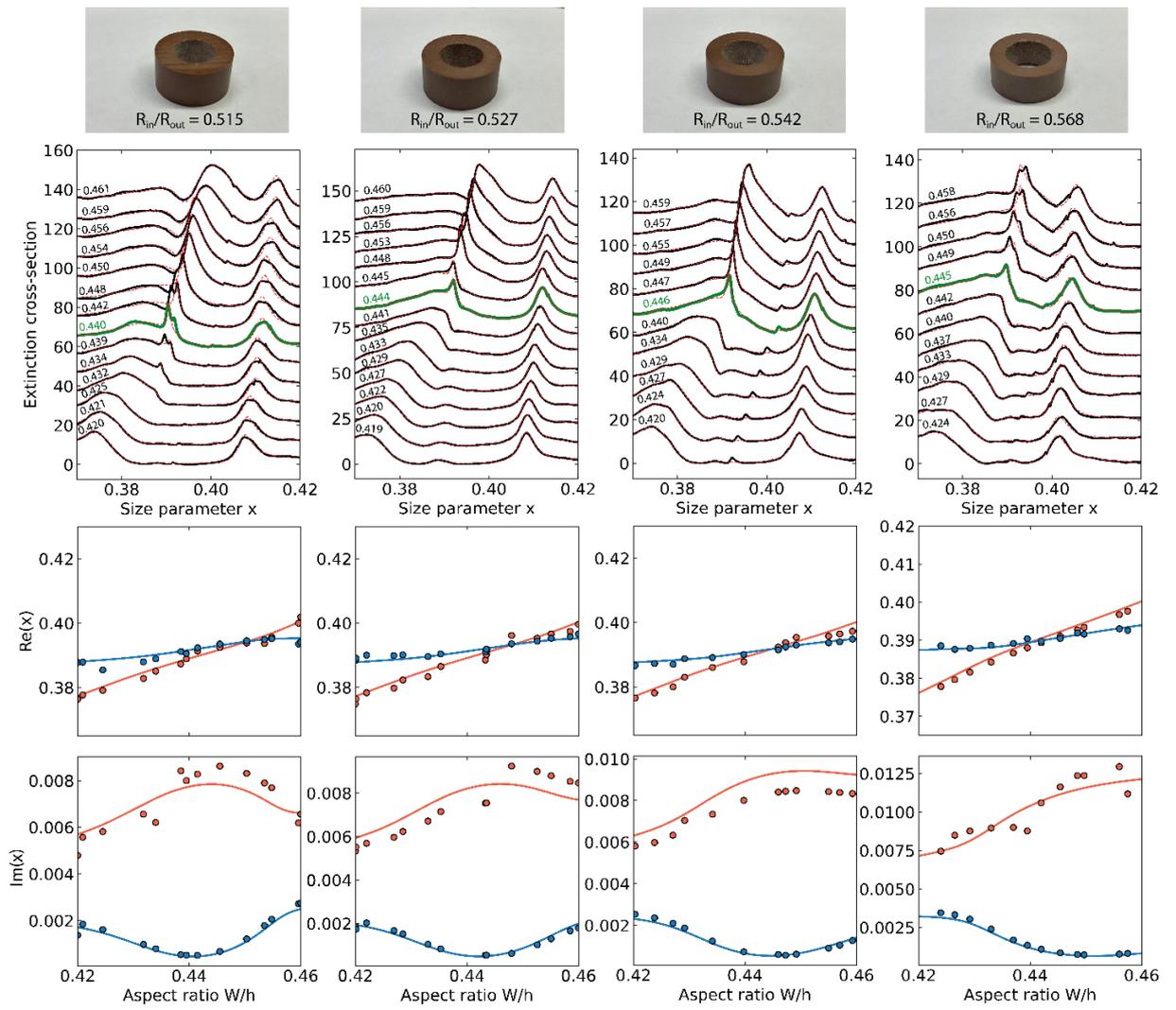

**Fig. S8** – Fitting of the experimental spectrum (black solid line) and eigenfrequencies vs W/h. Range of $0.5 < R_{in}/R_{out} < 0.57$.

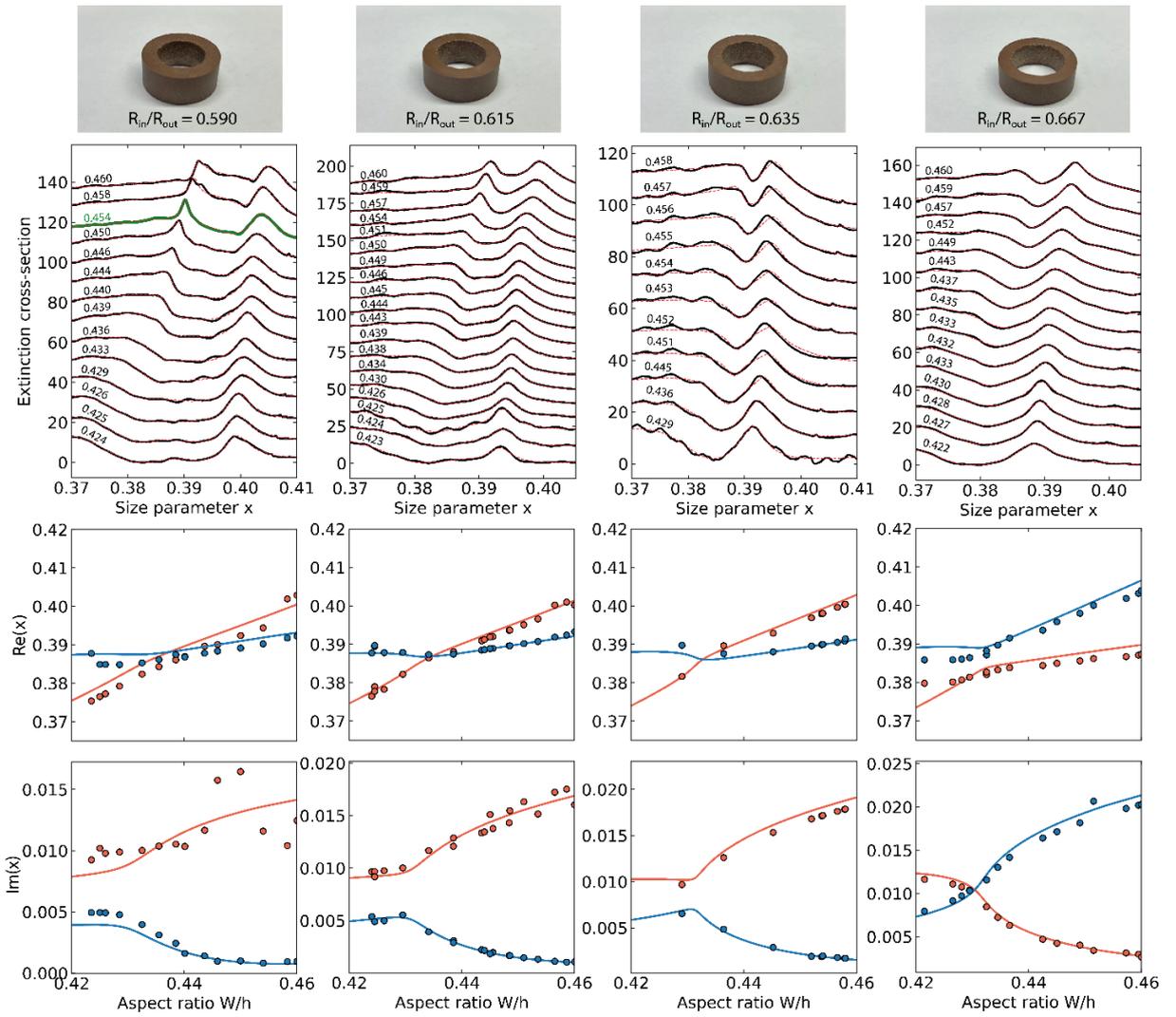

**Fig. S9** – Fitting of the experimental spectrum (black solid line) and eigenfrequencies vs W/h. Range of $0.58<R_{in}/R_{out}<0.7$.

In each column corresponding to a fixed $R_{in}/R_{out}$, a photograph of the sample is presented at the top, and below are the experimental extinction spectra and their fitting (red dotted line) and the experimental eigenvalues (represented by dots) and a comparison with the calculations of the eigenfrequencies (solid lines). The extinction spectra corresponding to q-BIC are highlighted in green, and the closest spectra corresponding to EP are highlighted in purple.

**References:**
[1] Lalanne, P., Yan, W., Kevin, V., Sauvan, C. & Hugonin, J.-P. Light interaction with photonic and plasmonic resonances, *Laser & Photonics Reviews* **12**, 1700113 (2018).